\documentstyle[epsfig]{mn}
\newif\ifAMStwofonts
\AMStwofontstrue

\def\be{\begin{equation}}
\def\ee{\end{equation}}

\def\Msun{{M_\odot}}
\def\Zsun{{Z_\odot}}

\def\gsim{\lower.5ex\hbox{\gtsima}}
\def\lsim{\lower.5ex\hbox{\ltsima}}
\def\gtsima{$\; \buildrel > \over \sim \;$}
\def\ltsima{$\; \buildrel < \over \sim \;$}
\def\prosima{$\; \buildrel \propto \over \sim \;$}
\def\gsim{\lower.5ex\hbox{\gtsima}}
\def\lsim{\lower.5ex\hbox{\ltsima}}
\def\simgt{\lower.5ex\hbox{\gtsima}}
\def\simlt{\lower.5ex\hbox{\ltsima}}
\def\simpr{\lower.5ex\hbox{\prosima}}
\def\ie{{\frenchspacing\it i.e. }}

\ifoldfss
  \ifCUPmtlplainloaded \else
    \NewTextAlphabet{textbfit} {cmbxti10} {}
    \NewTextAlphabet{textbfss} {cmssbx10} {}
    \NewMathAlphabet{mathbfit} {cmbxti10} {} 
    \NewMathAlphabet{mathbfss} {cmssbx10} {} 
  \fi
  \ifAMStwofonts
    \ifCUPmtlplainloaded \else
      \NewSymbolFont{upmath} {eurm10}
      \NewSymbolFont{AMSa} {msam10}
      \NewMathSymbol{\upi}     {0}{upmath}{19}
      \NewMathSymbol{\umu}     {0}{upmath}{16}
      \NewMathSymbol{\upartial}{0}{upmath}{40}
      \NewMathSymbol{\leqslant}{3}{AMSa}{36}
      \NewMathSymbol{\geqslant}{3}{AMSa}{3E}

      \let\leq=\leqslant \let\le=\leqslant
       \let\ge=\geqslant
    \fi
  \fi
\fi 

\ifnfssone
  \newmathalphabet{\mathit}
  \addtoversion{normal}{\mathit}{cmr}{m}{it}
  \addtoversion{bold}{\mathit}{cmr}{bx}{it}
  \newmathalphabet{\mathbfit} 
  \addtoversion{normal}{\mathbfit}{cmr}{bx}{it}
  \addtoversion{bold}{\mathbfit}{cmr}{bx}{it}
  \newmathalphabet{\mathbfss} 
  \addtoversion{normal}{\mathbfss}{cmss}{bx}{n}
  \addtoversion{bold}{\mathbfss}{cmss}{bx}{n}
  \ifAMStwofonts
    \ifCUPmtlplainloaded \else
      \UseAMStwoboldmath
      \makeatletter
      \new@mathgroup\upmath@group
      \define@mathgroup\mv@normal\upmath@group{eur}{m}{n}
      \define@mathgroup\mv@bold\upmath@group{eur}{b}{n}
      \edef\UPM{\hexnumber\upmath@group}
      \new@mathgroup\amsa@group
      \define@mathgroup\mv@normal\amsa@group{msa}{m}{n}
      \define@mathgroup\mv@bold\amsa@group{msa}{m}{n}
      \edef\AMSa{\hexnumber\amsa@group}
      \makeatother
      \mathchardef\upi="0\UPM19
      \mathchardef\umu="0\UPM16
      \mathchardef\upartial="0\UPM40
      \mathchardef\leqslant="3\AMSa36
      \mathchardef\geqslant="3\AMSa3E

      \let\leq=\leqslant \let\le=\leqslant
       \let\ge=\geqslant
    \fi
  \fi
\fi 

\ifnfsstwo
  \DeclareMathAlphabet{\mathbfit}{OT1}{cmr}{bx}{it}
  \SetMathAlphabet\mathbfit{bold}{OT1}{cmr}{bx}{it}
  \DeclareMathAlphabet{\mathbfss}{OT1}{cmss}{bx}{n}
  \SetMathAlphabet\mathbfss{bold}{OT1}{cmss}{bx}{n}
  \ifAMStwofonts
    \ifCUPmtlplainloaded \else
      \DeclareSymbolFont{UPM}{U}{eur}{m}{n}
      \SetSymbolFont{UPM}{bold}{U}{eur}{b}{n}
      \DeclareSymbolFont{AMSa}{U}{msa}{m}{n}
      \DeclareMathSymbol{\upi}{0}{UPM}{"19}
      \DeclareMathSymbol{\umu}{0}{UPM}{"16}
      \DeclareMathSymbol{\upartial}{0}{UPM}{"40}
      \DeclareMathSymbol{\leqslant}{3}{AMSa}{"36}
      \DeclareMathSymbol{\geqslant}{3}{AMSa}{"3E}

      \let\leq=\leqslant \let\le=\leqslant
       \let\ge=\geqslant
    \fi
  \fi
\fi 

\ifCUPmtlplainloaded \else
  \ifAMStwofonts \else 
    \def\upi{\pi}
    \def\umu{\mu}
    \def\upartial{\partial}
  \fi
\fi

\title[Constraints on the IMF of the first stars]{Constraints on the IMF of the first stars}

\author[Raffaella Schneider, Ruben Salvaterra, Andrea Ferrara \& Benedetta Ciardi]{Raffaella Schneider$^{1,2}$, Ruben Salvaterra$^3$, Andrea Ferrara$^4$ \& Benedetta Ciardi$^5$\\
$1$ Centro Enrico Fermi, Via Panisperna 98/A, 00184 Roma, Italy\\
$2$ Osservatorio Astrofisico di Arcetri, Largo Enrico Fermi 5, 50125 Firenze, Italy\\
$3$ Dipartimento di Fisica e Matematica, Universit\`a dell'Insubria, Via Valleggio 11, 22100, Como, Italy \\
$4$ SISSA/Internation School for Advanced Studies, Via Beirut 4, 34100 Trieste, Italy\\ 
$5$ Max-Planck-Institut f{\"u}r Astrophysik, 85741 Garching, Germany}

\date{20 Oct 2005}

\pagerange{\pageref{firstpage}--\pageref{lastpage}}
\pubyear{2005}

\begin{document}

\maketitle
\label{firstpage}

\begin{abstract}
Motivated by theoretical predictions that first stars were 
predominantly very massive, we investigate the physics of the transition from an early
epoch dominated by massive Pop~III stars to a later epoch dominated 
by familiar low-mass Pop~II/I stars by means of a numerically-generated catalogue 
of dark matter halos coupled with a self-consistent treatment of chemical and 
radiative feedback. Depending on the strength of the chemical feedback, Pop~III stars
can contribute a substantial fraction (several percent) of the cosmic star formation activity 
even at moderate redshifts, $z\approx 5$. We find that 
the three $z \approx 10$ sources tentatively detected in NICMOS UDFs should be powered by Pop~III stars, if
these are massive; however, this scenario fails to reproduce the derived WMAP electron scattering optical depth.
Instead, both the UDFs and WMAP constraints can be fulfilled if stars at any time form with
a more standard, slightly top-heavy, Larson IMF in the range $1 M_\odot \la M_\star \la 100 M_\odot$.  
\end{abstract}


\section{Introduction}

In the last few years, the universality of the stellar initial mass function (IMF)
has been questioned by the results of theoretical studies, which consistently predict
that the first stars (hereafter Pop III stars) had characteristic masses of 100-600 $\Msun$ 
(Omukai \& Nishi 1998; Abel, Bryan \& Norman 2002; Bromm, Coppi \& Larson 2002; 
Omukai \& Palla 2003), about 100 times more massive than those observed today (Larson 2003).

Indeed, the physical conditions in primordial star-forming regions appear to systematically
favor the formation of very massive stars. 
In particular, (i) the fragmentation scale of metal-free clouds is typically 
$10^{3}\;\Msun$ (Abel, Bryan \& Norman 2002; Bromm, Coppi \& Larson 2002)
(ii) because of the absence of dust grains the radiative feedback from 
the forming star is not strong enough to halt further gas accretion 
(Omukai \& Palla 2003).
(iii) since the accretion rate is as large as $10^{-3}$-$10^{-2}\;\Msun \rm{yr}^{-1}$,
the star grows up to $\gsim 100\,\Msun$ within its lifetime  
(Stahler, Palla \& Salpeter 1986; Omukai \& Nishi 1998; Ripamonti et al. 2002).    

In spite of the significant progresses made by numerical simulations
and semi-analytic models, many aspects of primordial star formation 
remain to be fully understood, the more important of which are the radiative
feedback effects which operate during protostellar accretion 
and which are likely to ultimately set the final stellar
mass. By modeling the structure of the accretion flow and the
evolution of the protostar, Tan \& McKee (2004) have recently shown
that radiative feedback becomes dynamically significant at 
protostellar masses $\approx 30\,\Msun$, and are likely to constrain the
mass of first stars in the range $100 - 300\,\Msun$.

On the other hand, observations of present-day stellar populations show that stars form 
according to a Salpeter IMF with a characteristic mass of 1 $\Msun$, below which the IMF 
flattens. Thus, unless the current picture of primordial star
formation is lacking in some fundamental physical process, a 
transition in the modes of star formation  must have
occurred during cosmic evolution. 

The physical processes responsible for this transition,
as well as its cosmological consequences, are currently the subject of systematic 
investigation. In particular, following the early study by Yoshii \& Sabano (1980), it
has been shown that a  key element driving this transition is 
the metallicity of the star-forming gas (Omukai 2000; Bromm et al. 2001; 
Schneider et al. 2002, 2003). The fragmentation properties of the collapsing clouds change 
as the mean metallicity of the gas increases above a critical threshold, 
$Z_{\rm cr} = 10^{-5 \pm 1} \; \Zsun$ (Schneider et al. 2002, 2003). 
The fragmentation of clouds with $Z<Z_{\rm cr}$ proceeds only down to
relatively massive ($>100\;\Msun$) cores, 
whereas in clouds with $Z>Z_{\rm cr}$ lower mass fragments can be formed. 
Within the critical metallicity range, low-mass cores can form if a sufficient amount of metals
are depleted onto solid dust grains, which provide an efficient cooling channel in the high
density regime (Schneider et al. 2003; Omukai et al. 2005).  

On cosmic scales, the transition between an epoch dominated by the
formation of massive Pop III stars to an epoch dominated by the
formation of ordinary Pop II/I stars is 
controlled by the strength of the so-called {\it chemical feedback}: 
the explosions of the first massive supernovae starts
to metal-enrich the gas out of which subsequent stellar generations
form. In particular, in regions of the Universe where the metallicity 
exceeds $Z_{\rm cr}$, low-mass stars can form. Chemical feedback
self-propagates through the metal enrichment process and therefore 
depends on many poorly constrained parameters, such as the star
formation efficiency at high redshifts, the number of supernovae per
stellar mass formed, the efficiency of metal ejection and mixing. In
particular, the shape of the primordial
IMF governs the number of Pop III stars with masses 
in the pair-instability supernova range (PISN, Heger \& Woosley 2002)
140 $\Msun$ - 260 $\Msun$ (which contribute to the metal
enrichment) relative to that of Pop III stars which end up as
massive black holes (which instead do not contribute to the metal
enrichment as all the heavy elements are captured by the black hole).

Using a semi-analytic approach, Scannapieco, Schneider \& Ferrara
(2003) have modeled the probability that a collapsing halo has been
impacted by metal-enriched outflows from  neighboring
halos. Exploring different chemical feedback strengths, they concluded 
that the transition between the Pop III and Pop II/I dominated 
star formation epochs is essentially controlled by the spatial distribution of
metals. Because metal enrichment is generally inefficient (the filling
factor of metals is always found to be $< 1$), the transition epoch
is extended in time, with coeval Pop III and Pop II/I star formation
episodes, and Pop III stars still contributing to the global star
formation rate at redshifts $z \lsim 5$. These findings have very
important implications for the detection of Pop III stars through
their strong Ly$\alpha$ emission (Scannapieco et al. 2003) as
well as the corresponding pair-instability supernova rate (Scannapieco
et al. 2005). 

In this work, we investigate the cosmic transition between Pop III and
Pop II/I stars following a complementary approach. We use the code
PINOCCHIO\footnote{http://www.daut.univ.trieste.it/pinocchio.}
 developed by Monaco, Theuns \& Taffoni (2002) and Taffoni,
Monaco \& Theuns (2002) to produce random catalogues of dark matter
halos at different redshifts. We then propagate metal enrichment
along the hierarchy of mergers of star-forming galaxies to predict
the number and mass distribution of metal-free objects, which can host
Pop III stars, at each redshift. We then couple these findings to the
star formation history (of Pop III and Pop II/I stars) and to the
history of cosmic reionization and quantify their interplay through
radiative feedback effects which can contribute to the extinction of Pop III stars. 

A similar approach has been recently undertaken by Furlanetto \& Loeb (2005) 
with the aim of assessing the plausibility of the double reionization
scenario suggested by Cen (2003) as a way to meet both the constraints
on the gas neutral fraction implied by observations of Gunn-Peterson 
troughs in the spectra of distant quasars (Fan et al. 2002, 2003) and
on the gas ionized fraction derived by the high optical depth to
Thomson scattering measured by the WMAP satellite (Spergel et
al. 2003; Kogut et al. 2003). 

Here we instead focus on the extinction epoch of Pop III stars which
we believe to be regulated by feedback both of radiative and chemical
types. We investigate the implications that complementary observations
on galaxy number counts at redshifts $z \approx 10$ (Bouwens et
al. 2005) and on the history of cosmic reionization (Kogut et al. 2003) 
set on the transition epoch between Pop III and Pop II/I stars and on 
the nature of Pop III stars as sources of UV photons contributing 
to cosmic reionization.

In particular, the detection of galaxies at the highest redshifts is 
challenging because of the redshifting of UV light into the infrared 
and the well-known limitations of current IR instruments. 
In spite of these difficulties, the search for high redshift galaxies, 
based on the dropout technique, has provided the first tentative detection 
of galaxies at redshifts beyond $z\approx 7$ (Bouwens et al. 
2004), stimulating a re-analysis of the small amount of deep IR data.
Bouwens et al. (2005) looked at the prevalence of galaxies at $z\approx 8-12$ 
by applying the dropout technique to the wide variety of deep fields
that have been imaged with the Near Infrared Camera and Multi-Object 
Spectrometer (NICMOS) on board of the Hubble Space Telescope. 
Using an appropriate selection criterion for high-$z$
sources, Bouwens et al. (2005) concluded that the number 
of $z\simeq 10$ sources must be three or fewer. 
When combined with the high optical depth to Thomson scattering 
inferred from the {\it Wilkinson Microwave Anisotropy Probe} (WMAP) 
satellite (Kogut et al 2003), these data already provide a demanding
benchmark for theoretical models.  
 
The paper is organized as follows. In Section~2 we describe the
different feedback effects that control the transition from a 
Pop~III to a Pop~II dominated star formation epoch. We illustrate 
the adopted chemical feedback model, predicting the number and
mass function of halos which can host Pop III stars at different
redshifts.  In Section~3, we couple these predictions to a star formation
and reionization model.
In Section~4 we
summarize the observations of galaxy counts at redshifts $z=10$ and
discuss how these can be used to set important constraints on the model. 
Finally, in Section~5 we present the main results and in Section~6 we
discuss their implications.  

Throughout the paper, we adopt a $\Lambda$CDM cosmological model with 
parameters $\Omega_M = 0.3$, $\Omega_{\Lambda} = 0.7$, $h=0.7$,
$\Omega_B=0.04$, $n=1$ and $\sigma_8=0.9$, which lie within the 
experimental errorbars of WMAP experiment (Spergel et al. 2003). 
We also use the AB magnitude 
system\footnote{AB magnitudes are defined
as AB$=-2.5\log (F_{\nu_{0}}) -48.6$, where $F_{\nu_{0}}$ is the spectral energy density
within a given passband in units of erg s$^{-1}$ cm$^{-2}$ Hz$^{-1}$}.

\section{Feedback-regulated transition}

At least three types of feedback effects can be relevant to the extinction of
the first massive stars: 
(i) radiative feedback caused by H$_2$ photo-dissociating radiation emitted 
by the first stellar sources, which can inhibit gas cooling and star formation 
inside small halos 
(ii) chemical feedback due to the release of metals and dust in the first 
Pop III supernova explosions, which pollute the gas out of which subsequent 
generations of Pop II stars form 
(iii) radiative feedback due to photo-heating filtering caused by 
cosmic reionization, \ie the increased temperature of the cosmic gas 
suppresses the formation of galaxies below the Jeans mass. 
A thorough discussion of each of these processes is beyond the scope of this paper and 
can be found in the recent extensive review by Ciardi \& Ferrara (2005).   

As a first step, we build up the hierarchy of mergers of dark halos hosting 
the star formation sites. We use PINOCCHIO code (Monaco, Theuns \& Taffoni 2002; 
Taffoni, Monaco \& Theuns 2002) to generate random catalogues of dark matter halos 
in a hierarchical universe. 

\begin{figure}
\centerline{\epsfig{figure=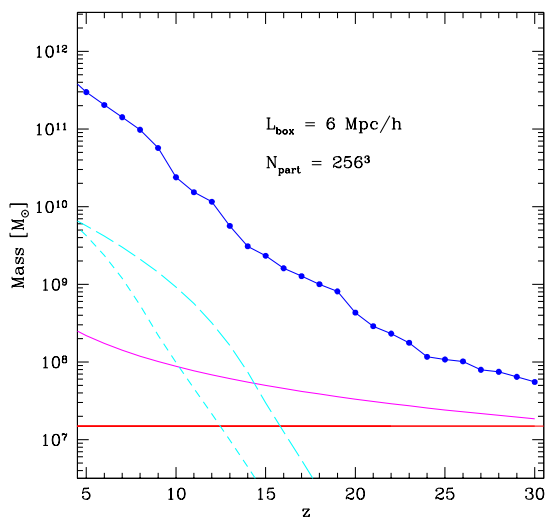, height=8.0cm}}
\caption{\footnotesize Properties of the reference PINOCCHIO run. The
  horizontal line shows the mass of the smallest resolved halo 
  and the line with points indicates the mass of the largest halo in
  the box at each simulation output. The solid line shows the evolution of
  the minimum halo mass allowed to form stars,  corresponding to a virial temperature $T_{vir} = 10^4$~K. 
  The short- and long-dashed lines represent the
  evolution of the filtering mass for two assumed reionization
  histories where reionization is complete at redshifts $\approx 10$ and
  $\approx 16$, respectively (see Section 2.2).}
\end{figure}

Our reference run has $256^3$ particles within a volume of comoving side 6 Mpc~$h^{-1}$
so that the smallest resolved dark matter halos have total mass $1.5 \times 10^7 \Msun$ 
(in the merger trees only halos with at least 10 particles are considered). 
Fig.~1 summarizes the details of our reference PINOCCHIO run and shows the redshift evolution of the 
largest and smallest mass present in the simulation box and of the minimum halo mass that is allowed to form stars. 
The latter corresponds, at each redshift, to halos with virial temperatures $T_{vir}=10^4~$K, 
\be
M_{min} = 10^8 \Msun \left(\frac{1+z}{10}\right)^{-3/2} 
\label{eq:min}
\ee
\noindent
so that 
stars do not form in halos which rely only on H$_2$ cooling 
(radiative feedback).  
The chosen resolution ensures that we can follow the formation 
of stars in the lowest mass halos at the highest redshift. 
Due to radiative feedback, the first star-forming halos have
a total mass $\approx 2 \times 10^7 \Msun$ and start to form at 
$z \approx 30$. This resolution is appropriate to follow
the transition from a Pop~III to a Pop~II~star formation mode around
redshift $z \approx 10$, which is the main scope of this analysis. 

In Fig.~2 we compare the resulting mass function of existing halos at redshifts 
30, 20, 10 and 5 in the simulation box, to the analytic predictions of
Press \& Schechter (1974) and Sheth \& Tormen (1999). The errorbars
are Poisson errors on the counts in each mass-bin.
The agreement is very good down to redshift $z \lsim 5$ showing that
the chosen resolution ensures completeness in the simulation volume
in the redshift range relevant to our analysis.

\begin{figure}
\centerline{\epsfig{figure=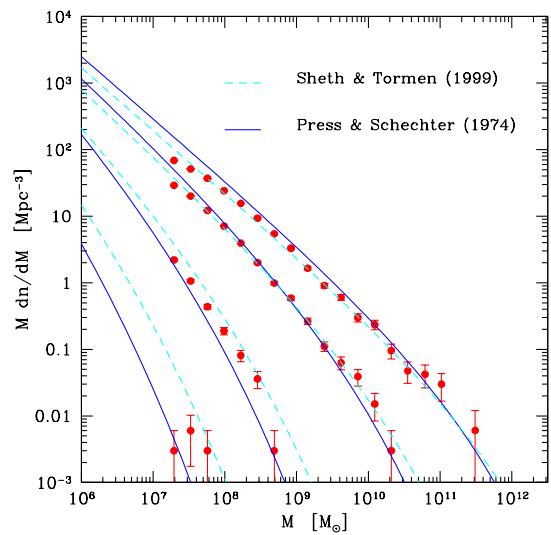, height=8.0cm}}
\caption{\footnotesize Comparison between the predicted mass
function of halos within the simulation box (points with errorbars)
and the analytic predictions of Press \& Schechter (1974, solid lines) and Sheth \&
Tormen (1999 dashed) at redshifts 30, 20, 10 and 5 (from left to
right). Data at redshift 5 have been shifted upward by a factor 0.3
for clarity. The errorbars are Poisson errors on the counts in each
mass bin.} 
\end{figure}

\subsection{Chemical feedback}
\label{sec:chem}

At each redshift, existing halos which are allowed to form stars 
($T_{vir}\ge 10^4$~K) are classified as hosting Pop II (Pop III) 
stars depending on whether the halo itself or any of its progenitors 
have (have not) already experienced an episode of star formation, leading
to the formation of SN progenitors which then metal-enrich the gas.
Note that even a single 200 $\Msun$ PISN can enrich the gas within a halo
with mass $M_{min}$ to a metallicity level well above the critical value as,

\be
Z = \frac{0.45 M_*}{(\Omega_B/\Omega_M)M_{min}} \approx
0.02 Z_\odot \left (\frac{1+z}{10} \right)^{3/2}
\ee 

where, following Heger \& Woosley (2002), we have assumed that the mass of metals 
released in a PISN explosion is 45\% of the initial progenitor mass.

This way of taking into account chemical feedback is very conservative as 
(i) we are assuming that all metals synthesized during a star formation 
episode remain 
confined (or are re-accreted) within the host halo and/or 
(ii) even if a fraction of metals is ejected out of the host halo, 
the metals that are kept within the potential well are sufficient to pollute 
the gas within the host system or within any unpolluted halo that
the system will later merge with, to a metallicity level larger than $Z_{\rm cr}$. 
On the other hand, we are neglecting metal pollution from neighboring 
galaxies through enriched-winds (Scannapieco et al. 2003; Sigward, Ferrara \& Scannapieco 2005). 

Fig.~3 shows the fraction of halos which can host 
Pop III stars (hereafter Pop III halos) 
at different redshifts as a function of their mass.
The three different curves correspond to different assumptions on the
shape of the primordial IMF. In particular, we randomly classify
Pop~III halos as SN-forming or BH-forming depending on whether
they host stars with masses in the PISN range or outside this range.
SN-forming halos are metal-enriched and pollute their
descendants. BH-forming halos are not enriched in metals and therefore
do not propagate chemical feedback along their subsequent hierarchy of
mergers. 
The fraction of SN-forming halos, $f_{\rm sn}$, 
is therefore a free parameter that controls the strength of chemical
feedback.  

The three curves shown in each panel of Fig.~3 correspond (from bottom
 to top) to $f_{\rm sn}=1, 0.5, 0.1$. 
Each curve represents an average over 50 realizations of the random
classification procedure and the errors are Poisson errors on the 
counts in each mass bin.

When $f_{\rm sn} = 1$, all Pop III halos are SN-forming and 
chemical feedback is strong. The total fraction of Pop III
halos decreases with time, being $\approx 0.25$ at $z=20$ and
$\approx 0.1$ at $z=5$; at all redshifts, Pop III stars are confined 
to form in the smallest halo mass bins (close to the minimum mass to form stars shown
as the vertical dotted line in each panel). For $f_{\rm sn} = 0.5$
(50\% of Pop III halos host SN), the number of Pop III halos at each
redshift is larger than the previous case ($\approx 0.5$ at $z=20$ and
$\approx 0.2$ at $z \approx 5$) but Pop III halos are still
confined to the smallest mass bins ($ \lsim 10^9 \Msun$). Finally,
when $f_{\rm sn} = 0.1$ the strength of chemical feedback is
significantly reduced (as only 10\% of Pop III halos at each redshift
host SN) and therefore the fraction of Pop III halos is dominant
at all redshifts, being $\approx 0.8$ at $z=20$ and $\approx 0.65$ at $z=5$,     
not negligible even in halos with the largest mass present
in the simulation box at each redshift.  Hereafter we only consider
$f_{\rm sn} = 0.1$ and $=1$ as representative of weak and strong 
chemical feedback models. 

Despite the strength of chemical feedback, Pop III halos
continue to form stars even at redshifts as small as $z \lsim 5$,
in agreement with the results of Scannapieco et al. (2003). Thus,
independently of whether it preferentially propagates through galaxy
mergers or through metal-enriched winds, chemical feedback always
leads to a smooth transition between a Pop III and a Pop II dominated 
star formation epoch. Note that the above results depend only on the
parameter $f_{\rm sn}$ and on the hierarchy of galaxy mergers, and are
totally independent of the star formation efficiency or of the value
of critical metallicity for the transition in fragmentation scales
of collapsing pre-stellar clouds.

\begin{figure}
\centerline{\epsfig{figure=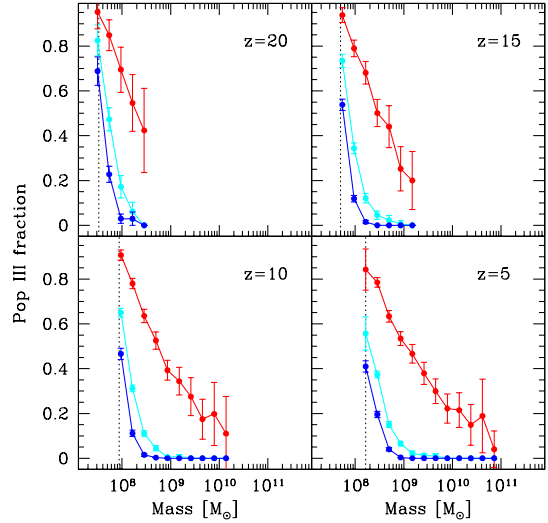, height=8.0cm}}
\caption{\footnotesize Fraction of star forming halos which, 
at a given redshift, hosts Pop III stars as a function of 
their total mass. The different panels correspond to 
different output redshifts, $z =20, 15, 10$ and 5. In each panel,
the three curves correspond to $f_{\rm sn} = 0.1, 0.5, 1$ (from
top to bottom) and the vertical dotted line shows the mass that
at each considered redshift corresponds to the minimum halo
mass to form stars, $M = M_{min}$ (see text).}
\end{figure}
\begin{table*}
\begin{center}
\caption{\footnotesize
Properties of the three models considered: $f_{sn}$ is the
strength of chemical feedback (the fraction of Pop~III halos
which hosts SN and therefore contributes to metal enrichment),
columns 3-5 (6-8) show the initial stellar metallicity, IMF type and 
Log of HI-ionizing photons per baryons in stars for Pop~III (Pop~II) stars.  
By Top-Heavy IMF we mean that only stars with masses $\ge 100~\Msun$
are considered. 
Finally, the last two columns show the minimum and maximum values for
($f_{esc}, f_{\star}$) allowed by the observed NICMOS-UDFs counts at
redshift $z=10$ (see text).}  
\begin{tabular}{ccc|c|c|ccc|c|c|ccc|c|cc|cc|}\hline
      &          &            &   Pop III  &              &   & Pop II   &    &   &  \\ \hline       
Model & $f_{sn}$ &$Z/Z_\odot$ &    IMF     & $N_{\gamma}$ &$Z/Z_\odot$ & IMF &
$N_{\gamma}$ & $[f_{esc},f_\star]_{min}$ & $[f_{esc},f_\star]_{max}$
\\ \hline 
 
 A    &   0.1 &  0 &  Top-Heavy  &  4.862  &  0.2 & Salpeter & 3.602 & $[0,0.9 \times 10^{-3}]$ & $[0.5, 1.65 \times 10^{-3}]$ \\    
 B    &   1   & 0 &  Top-Heavy  &  4.862  &  0.2 & Salpeter & 3.602 & $[0,1.15 \times 10^{-2}]$ & $[0.5, 2.15 \times 10^{-2}]$ \\
 C    &   1   & 0 &  Salpeter   &  4.235  &  0.2 & Salpeter & 3.602 & $[0,0.1]$ & $[0.5, 0.1]$ \\
\hline
\end{tabular}
\end{center}
\end{table*}

\subsection{Photoionization}

So far, we have taken into account radiative feedback only through the
choice of the minimum halo mass that can host stars. Namely, we have
assumed that mini-halos with virial temperatures less than $10^4~$K do 
not significantly contribute to the cosmic star formation history. 
However, radiative feedback does affect larger halos, particularly 
when the universe is approaching  the epoch of cosmic reionization. 
Indeed, the increase in temperature of cosmic gas during reionization 
will suppress the formation of small galaxies with mass below the Jeans mass. 
To quantify this effect, we follow the method proposed by Gnedin (2000) and adopt the filtering mass as the
characteristic scale below which the gas fraction in star forming objects, hence the fraction
of stellar mass, is reduced with respect to the universal value. 
In particular, before reionization, the average baryon fraction (the ratio of the mass in baryons over
the total halo mass) has the universal value $\Omega_B/\Omega_M$ while after reionization 
it can be approximated by the fitting formula 
(Gnedin 2000),

\be
f_b = \frac{\Omega_B/\Omega_M}{[1+(2^{1/3}-1)M_C/M]^3},
\label{eq:fil}
\ee

\noindent
where the characteristic mass, $M_C$, is the total mass of objects which, on average, retain 50\% of their
universal baryon mass, $0.5 (\Omega_B/\Omega_M) M_C$. 
This characteristic mass is very well approximated by the linear-theory filtering mass, $M_C \approx M_F$,
with
\be
M_F^{2/3} = \frac{3}{a} \int_0^a da^\prime M_J^{2/3}(T, a^\prime)[1-(\frac{a^\prime}{a})^{1/2}],
\ee

\noindent
where the Jeans mass is given by 

\be
M_J(T) = 2.56  \times 10^5 \Msun (\Omega_M h^2)^{-1/2}(T/\mu)^{3/2} (1+z)^{-3/2},
\ee

\noindent
with $T$ and $\mu$ the temperature and molecular weight of the cosmic gas. 

Here we assume the temperature of ionized regions to be $10^4$~K and
we modulate the increase in temperature of the cosmic gas due to reionization
with the volume filling factor of ionized regions at the corresponding redshift
$Q_{\rm HII}(z)$ (see Section~\ref{sec:reio}).  

Thus, the characteristic mass depends on the thermal history of the gas
and it accounts for the finite time required for pressure to influence 
the gas distribution. The redshift evolution of the filtering mass is 
shown in Fig.~1 for a model where reionization occurs at redshift
$z_{\rm rei} \approx 10$ (short-dashed line) and $\approx 16$ (long-dashed line).  
The filtering mass steeply increases with time and, as expected, its
value in the redshift range $5 \lsim z \lsim 20$ depends strongly on
the reionization history. At redshift $z \approx 10$, the filtering mass
is $\approx 10^8 \Msun$ for the late reionization case, and it is a
factor ten larger for the early reionization case. 

As can be inferred from Fig.~1 and ~3, following reionization, 
photoionization feedback can significantly
suppress the formation of Pop III stars since halos with masses
$10^8 \Msun \le M \le 10^9 \Msun$ have evaporated at least 50\% of
their mass.

\section{Reionization history}
\label{sec:reio}

Following Barkana \& Loeb (2001), we compute the redshift evolution of
the filling factor of HII regions as,
\be
Q_{\rm HII}(z) = \int_{z}^{\infty} dz^\prime
\left\vert\frac{dt}{dz}\right\vert \frac{1}{n_H^0}
\frac{dn_{\gamma}}{dt} \mbox{e}^{F(z^\prime,z)},
\label{eq:reio}
\ee

\noindent
where $n_H^0 = X_H n_B^0$ and $n_B^0$ are the present-day number densities of
hydrogen and baryons ($X_H=0.76$ is the hydrogen mass fraction), and
$dn_{\gamma}/dt$ is the production rate of ionizing photons that we
can express as,

\be
\frac{dn_{\gamma}}{dt} = f_{esc} N_{\gamma} \dot{\rho}_\star,
\ee

\noindent
where $f_{esc}$ is the escape fraction, \ie the fraction of 
ionizing photons that is able to escape the host galaxy,
$N_{\gamma}$ is the time-averaged number of ionizing photons 
emitted per unit stellar mass formed, and $\dot{\rho}_\star$
is the comoving star formation density,
\be
\dot{\rho}_\star = f_\star f_b \frac{d}{dt} \int_{M_{min}}^{\infty} dM M \frac{dn}{dM}(M,z) 
\ee 
where $f_\star$ is the star formation efficiency, $f_b$ is the baryon fraction given by
eq. (\ref{eq:fil}), $M_{min}$ is the minimum halo mass to form stars (see eq. \ref{eq:min})
and $n(M,z)$ is the number density of halos (Pop II and Pop III) at redshift $z$. 
The parameter $f_{esc}$ also affects the relative strength of the stellar continuum and the
Ly$\alpha$ line and nebular luminosity of each galaxy. The latter
contributions are particularly important for massive Pop~III stars.

Finally, the function $F(z',z)$ takes into account the
effect of recombinations. Assuming a time-independent
volume-averaged clumping factor $C$, common to all HII regions, 
we can write, 

\be
F(z^\prime,z) = -\frac{2}{3} \frac{\alpha_B n_H^0}{\sqrt{\Omega_M} H_0} C 
[f(z^\prime)-f(z)],
\ee

\noindent
and 

\be
f(z)=\sqrt{(1+z)^3 + \frac{1-\Omega_M}{\Omega_M}},
\ee

\noindent
where $\alpha_B = 2.6 \times 10^{-13} \mbox{cm}^3 \mbox{s}^{-1}$ is
the hydrogen recombination rate. 

For a given reionization history, it is straightforward to compute the optical
depth to electron scattering as,

\be
\tau_e(z)=\int_0^z dz^\prime \left\vert c \frac{dt}{dz^\prime}\right\vert \sigma_T n_e(z^\prime),
\ee

\noindent
where $\sigma_T = 6.65 \times 10^{-25} \mbox{cm}^{2}$ is the Thomson cross 
Section and $n_e(z)$ is the number density of free electrons at redshift $z$.
In this paper we only consider the evolution of the filling factor of HII regions ($Q_{\rm HII}$). 
However, for the purpose of computing the optical depth to 
electron scattering, we assume that $Q_{\rm HeII} = Q_{\rm HII}$ and we neglect 
the contribution of electrons coming from HeIII regions 
(see also Haiman \& Holder 2003). 
Under these assumptions we can write $n_e(z)=Q_{\rm HII}(z) n_B^0(1+z)^3$ 
where $Q_{\rm HII}(z)$ is given by eq.(6).

\section{Galaxy counts at redshift 10}

Bouwens et al. (2005) looked at the prevalence of galaxies at $z\approx 8-12$ 
by applying the dropout technique to the wide variety of deep
F110W- and F160W-band (hereafter $J_{110}$ and $H_{160}$, respectively) fields
that have been imaged with the Near Infrared Camera and Multi-Object 
Spectrometer (NICMOS). The principal data set is constituted by two
1.3 arcmin$^2$ deep NICMOS parallels taken with the Advanced Camera 
for Surveys (ACS) Hubble Ultra Deep Field (UDF; each has about $160$ orbits data).
Complementary shallower fields possessing similar $J_{110}+H_{160}$ imaging 
have been analyzed as well. The 5$\sigma$ limiting magnitude for the UDFs        
is  $\approx 28.6$ in $J_{110}$ and $\approx 28.5$ in $H_{160}$ 
($0.6^{\prime\prime}$ aperture). Since these limiting magnitudes are 
reached only in the deepest regions of the two NICMOS UDFs, we adopt the 
conservative limiting magnitude $H_{160}=28$ (Bouwens, private communication). 

The primary selection criterion for high-$z$
sources is $J_{110}-H_{160}>1.8$. Using this criterion, Bouwens et al. (2005) found
eleven sources. Eight of these are ruled out as credible $z\simeq 10$
sources, either as a result of detection ($>2\sigma$) blue-ward of $J_{110}$
or because of their colors red-ward of the break ($H_{160}-K \approx 1.5$). The nature
of the three remaining sources could not be assessed from the data, but
this number appears consistent with the expected contamination from 
low-redshift interlopers. Hence, Bouwens et al. (2005) concluded that
the actual number of $z\simeq 10$ sources in the NICMOS parallel fields 
must be three or fewer. 

Adopting the same selection criterion as in Bouwens et 
al. (2005), we compute the number of $z\approx
10$ galaxies detectable in the NICMOS UDFs predicted by our models.
Assuming that the three candidates selected by Bouwens et al. (2005)
are indeed $z\approx 10$ galaxies, we can set interesting constraints 
on the models. 
In particular, for a given chemical feedback strength $f_{\rm sn}$, we
compute the number density of Pop III and Pop II halos at each
redshift. We then assume a universal value for the star formation efficiency
$f_{\star}$ and escape fraction $f_{esc}$ of ionizing photons and we
characterize each stellar population (Pop III and Pop II/I stars) by
assuming a specific template emission spectrum. 
 
In the following, we consider two models, A and B, whose details
are given in Table~1. 
We assume two representative chemical feedback
strengths ($f_{\rm sn}$ = 0.1 for model A and $f_{\rm sn}$ = 1 for
model B). In both cases, 
the emission properties of Pop~II stars have been computed 
using a simple stellar population model taken from the GALAXEV library (Bruzual \& Charlot 2003).
Pop~II stars have an initial metallicity of $Z = 0.2\, Z_{\odot}$ and are assumed to form
according to a Salpeter IMF (with masses in the range 0.1 $\Msun$ - 100 $\Msun$). 
Pop~III stars are assumed to be very massive
(with masses larger than 100 $M_{\odot}$) and their emission properties are computed using the
stellar spectra of Schaerer (2002), including the nebular continuum emission found to be very
important for stars with strong ionizing fluxes\footnote{Since the emission spectrum of very massive
Z=0 stars is found to be independent of the stellar mass for stars
with masses $\ge 100~\Msun$ (Bromm, Kudritzki \& Loeb 2001; Schaerer
2002), we are not forced to make any assumption about the shape of the
IMF of Pop~III stars as long as these are assumed to be very
massive.}.

For each model, we then compute the number of sources per unit solid
angle with observed flux in the range $F_{\nu_0}$ and $F_{\nu_0}+dF_{\nu_0}$ as

\begin{equation}
\frac{dN}{d\Omega dF_{\nu_0}}(F_{\nu_0},z_{0})=\int^{\infty}_{z_{0}}
dz \left( \frac{dV_c}{dz d\Omega} \right) \, n(F_{\nu_0},z),
\label{eq:counts1}
\end{equation}

\noindent
where $dV/d\Omega dz$ is the comoving volume element per unit solid
angle and redshift, $n(F_{\nu_0},z)$ is the comoving number of objects at
redshift $z$ with observed flux in the range 
$[F_{\nu_0}, F_{\nu_0} + dF_{\nu_0}]$, given by

\begin{equation}
n(F_{\nu_0},z)=\int^{\infty}_z dz^\prime \, \frac{dM}{dF_{\nu_0}}
(F_{\nu_0},z,z^\prime) 
\, \frac{d^2 n}{dM dz^\prime}(M,z^\prime).
\label{eq:counts2}
\end{equation}

\noindent
Here $d^2 n/dMdz$ is the halo formation rate, and the factor $dM/dF_{\nu_0}$ 
converts the number density per unit mass in unit flux interval.

The average flux $F_{\nu_0}$ from a halo of mass $M$ at redshift $z$
that has formed at $z^\prime \ge z$ is,

\begin{equation}\label{eq:flux}
F_{\nu_0}=\frac{f_\star (\Omega_B/\Omega_M) M}{4 \pi
  \Delta\nu_0\, d_L(z)^2} \int^{\nu_{max}}_{\nu_{min}} 
d\nu \,\, l(\nu,t_{z,z^\prime}) \, \mbox{e}^{-\tau_{eff}(\nu_0,z_0,z)},
\label{eq:counts3}
\end{equation}

\noindent
where $l(\nu,t_{z,z^\prime})$
is the template luminosity per unit solar mass for a population of age 
$t_{z,z^\prime} $ (the time elapsed between the redshift $z^\prime$ and 
$z$), $d_L(z)$ is the luminosity distance, $\Delta\nu_0$ is the instrumental bandwidth, 
$\nu_{min}$ and $\nu_{max}$ are the restframe frequencies corresponding to the observed ones. 
Finally, $\tau_{eff}$ is the intergalactic medium (IGM) effective optical depth at $\nu_0$ between
redshift $z_0$ and $z$ (see Sect. 2.2 of Salvaterra \& Ferrara 2003)

\section{Results}

We compute the expected number of $z \approx 10$ galaxies detectable in the NICMOS UDFs
using eqs.~\ref{eq:counts1}-\ref{eq:counts3} and applying the same selection criterion of
Bouwens et al. (2005) ($J_{110}-H_{160}>1.8$, integrated up to the magnitude
limit of the survey, $H_{160}=28$). Assuming that only the three most
luminous sources should be above the detection threshold, we derive the
maximum allowed value for the star formation efficiency,
$f_\star$, corresponding to a given escape fraction of ionizing
photons. 

The results obtained for models A and B are shown in Fig.~4 in terms of
comoving star formation density in Pop~III and Pop~II stars
as a function of redshift. Model C will be discussed in section \ref{sec:reio}.

In each panel, vertically (horizontally)
shaded regions represent the range of star formation
histories for Pop~III (Pop~II) stars allowed by the NICMOS-UDFs counts when
a value $0 \le f_{esc} \le 0.5$ in assumed (see Table~1 for the corresponding
values of $f_\star$).
For both models and all the explored values of
$f_{esc}$, the three most luminous objects in NICMOS UDFs are always
represented by galaxies hosting Pop~III stars, as their strong
Ly$\alpha$ and nebular emission (Schaerer 2002) is dominant. As
a consequence, $f_\star$ is restricted to relatively small values, which
increase with increasing $f_{esc}$.
Model B, being characterized by
the strongest chemical feedback, and therefore by the smallest fraction of
halos hosting Pop~III stars at redshift $\le 10$ (see also Fig.~3), has  
larger allowed star formation efficiencies than model A. 

\begin{figure}
\centerline{\epsfig{figure=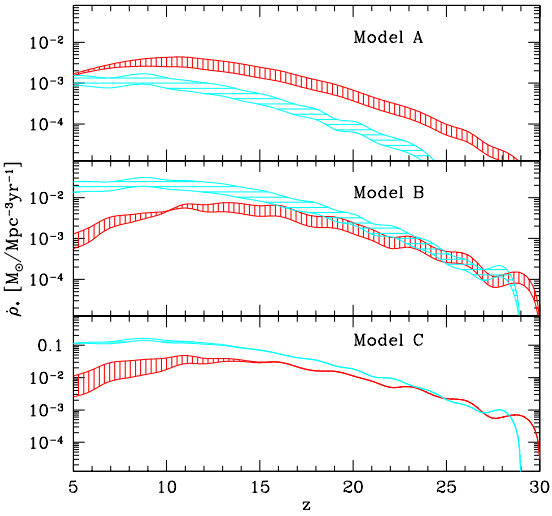, height=8.0cm}}
\caption{\footnotesize Star formation densities as a function of
  redshift in model A (upper panel), model B (middle panel) and 
model C (lower panel, see
  Table~1). The vertically (horizontally) shaded regions represent the
  range of star formation densities in Pop~III (Pop~II) stars allowed 
  by the observed NICMOS-UDFs counts at redshift $z = 10$ (see text).
  For all models, the IGM clumping factor is assumed to be $C=10$.}
\end{figure}

Fig.~4 shows that in model A chemical feedback is very inefficient 
($f_{sn}=0.1$) and 
Pop~III stars dominate the cosmic star formation history down to
redshift $z \le 5$. Conversely, model B shows that when chemical
feedback is maximally efficient, the star formation density in Pop~III
stars is sub-dominant at all but the highest redshifts, but Pop~III stars
still make a significant contribution to the cosmic star formation
history down to redshift 5. 

Because of the moderate star formation efficiencies implied by
the observed NICMOS UDFs counts, radiative feedback is negligible
for the most massive galaxies. To evaluate its impact on the predicted
star formation densities shown in Fig.~4, we have assumed a clumping factor 
$C=10$. For $0 \leq f_{esc} \leq 0.5$, the corresponding reionization redshifts 
vary in the range $0 \leq z_{\rm rei} \leq 10$ for model A, and in the range 
$0 \leq z_{\rm rei} \leq 14$ for model B.

In model A, radiative feedback effect on the overall Pop~III star formation density manifests
itself as a progressive decrement of the upper limit (which
corresponds to the largest $f_{esc}$ and $f_\star$) of the shaded
region for $z \lsim 10$. In model B, where the star
formation efficiencies implied by the observed counts are larger and,
at each redshift, Pop~III stars are confined to form only in the
smallest halos, radiative feedback is more significant; around redshift
10, the lines corresponding to the maximum and minimum Pop~III star formation
efficiencies cross each other (the highest the star formation
efficiency and escape fraction, the strongest the radiative feedback). 

Overall, the observed galaxy counts at redshift $z \approx 10$ favor a
cosmic star formation history where the transition from Pop~III to
Pop~II stars is governed by strong chemical feedback, with all 
Pop~III halos efficiently propagating metal enrichment to their  
descendants along the hierarchy of galaxy mergers. In fact, in
less efficient chemical feedback scenarios ($f_{sn} = 0.1$), the predicted
cosmic star formation density is dominated by Pop III stars down to 
redshift $z \approx 5$.

\subsection{Implications for reionization}
\label{sec:reio}

Stronger constraints on models A and B come from their corresponding 
reionization histories. Figs.~5 and ~6 show isocontours of
redshifts of reionization as a function of the clumping factor and 
the escape fraction (the star formation efficiency is set 
by the observed counts at $z \approx 10$). We define the redshift of
reionization, $z_{rei}$, as the redshift when the filling factor of 
ionized regions, given by eq.~(\ref{eq:reio}) is $Q_{\rm HII}(z_{\rm rei})=1$. 
The shaded regions indicate the
parameter space which leads to an optical depth to electron scattering
within the range $\tau_e = 0.16 \pm 0.04$ observed by the WMAP
satellite (Spergel et al. 2003; Kogut et al. 2003), with the dashed
line corresponding to the central value. 

\begin{figure}
\centerline{\epsfig{figure=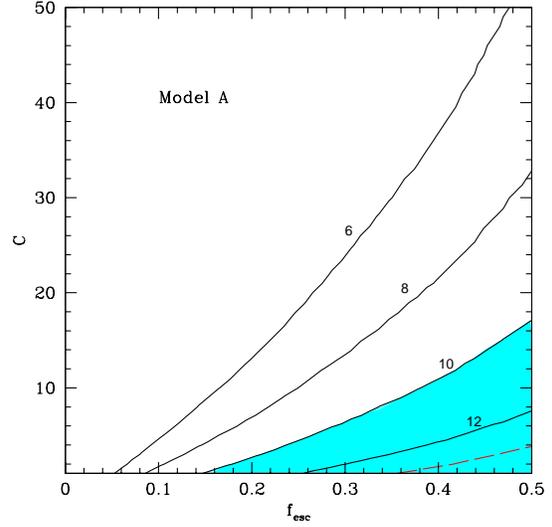, height=8.0cm}}
\caption{\footnotesize Isocontours corresponding to reionization
  redshifts 6, 8, 10, and 12 for model A in the plane ($C, f_{esc}$), 
where $C$ is the clumping factor and $f_{esc}$ the escape fraction. 
The dashed region indicates the parameter space which leads to
an optical depth to electron scattering within the range $\tau_e =
0.16 \pm 0.04$ observed by WMAP. The dashed line corresponds to the
central value.}
\end{figure}
\begin{figure}
\centerline{\epsfig{figure=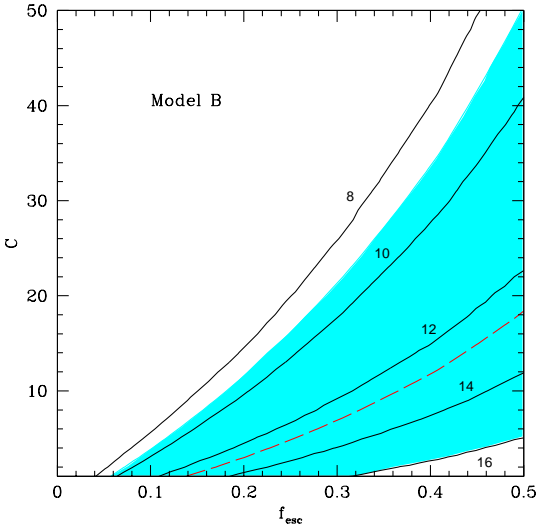, height=8.0cm}}
\caption{\footnotesize Same as Fig.~5 but for model B and reionization
  redshifts 8, 10, 12, 14 and 16.}
\end{figure}

As it can be inferred from the figures, both models lead to reionization
histories which are in poor agreement with observations. Indeed,  
large escape fractions and moderate clumping factors are required to 
reproduce the observed range of optical depths.
Independent observations of the escape fraction give, with few exceptions, values
in the range $f_{esc} \le 0.15$ (Ciardi \& Ferrara 2005 and references
therein) whereas numerical simulations (Gnedin \& Ostriker 1997;
Iliev, Scannapieco \& Shapiro 2005) 
and semi-analytic models (Madau, Haardt \& Rees 1999) predict clumping
factors in the range $1 < C \le 30$. 

Fig.~5 shows that in model A, even assuming an homogeneous IGM
($C=1$), the escape fraction has to be $f_{esc}>0.15$ to be 
marginally consistent with the observations. In model B, 
the agreement improves but still requires extreme assumptions 
on either the escape fraction or the clumping factor: 
when $f_{esc} \le 0.15$, the predicted optical depth lies 
within -1$\sigma$ error from the observed value only if $C < 8$.

Our analysis suggests that it is very hard to reconcile the observed
number counts at redshift $z \approx 10$ with the reionization history
inferred by WMAP observations if Pop~III stars are assumed to be all
very massive, with masses $\ge 100 \,\Msun$. This result is somehow 
counter-intuitive because metal-free very massive stars have always 
believed to play an important role in an early epoch of cosmic 
reionization (Cen 2003; Haiman \& Holder 2003; Sokasian et al. 2004,
2004; Venkatesan, Tumlinson \& Shull 2003; Wyithe \& Loeb 2003; 
but see also Ciardi, Ferrara \& White 2003, and Ricotti \& Ostriker
2004 for alternative interpretations).  

Still, due to their strong Ly$\alpha$ and nebular luminosity,
Pop~III stars with masses $\ge 100 \, \Msun$ always dominate the
predicted galaxy number counts at $z \approx 10$, even assuming the
strongest chemical feedback model. The paucity of candidate objects
observed in NICMOS UDFs implies that star formation in halos with
$T_{vir} \ge 10^4$~K must have occurred very inefficiently. 
Even if all three objects identified by Bouwens et al. (2005) 
are truly $z \approx 10$ sources, the contribution of Pop~III stars 
to cosmic reionization is strongly constrained. 

The constraints are particularly severe when chemical
feedback is less efficient (model A), because Pop~III stars can 
form at redshift $z \approx 10$ in relatively large, and therefore 
luminous, galaxies. Even if chemical feedback is assumed to be maximally
efficient and Pop~III stars are confined to form in the smallest
halos, their contribution to cosmic reionization is
in agreement with observations only for moderate ($C<8$) clumping
factors.

Unless an additional type of feedback comes into play, the transition
between Pop~III to Pop~II halos, regulated by chemical and radiative 
feedback, is very smooth. Thus, to relax the 
severe upper limits placed on the star formation efficiencies 
by NICMOS UDFs counts, the emission properties of Pop~III stars have
to be modified. 

In particular, if Pop~III stars are assumed to form according to a 
Salpeter IMF with masses in the range $1~\Msun \le M_\star \le
100~\Msun$, their Ly$\alpha$ and nebular emission is strongly
suppressed. In this limit, (see model C in Table~1), all Pop~III 
halos contribute to chemical feedback, $f_{sn}=1$, because each 
burst of star formation leads to the formation of SN progenitors 
and NICMOS UDFs counts at $z \approx 10$ are dominated by Pop~II
galaxies. 
As a consequence, the upper limit on the star formation efficiency 
set by the observed counts is $f_\star = 0.1$, much higher than those
found for models A and B and independent of the escape
fraction. 

The predicted evolution of the star formation density in Pop II and Pop III stars
is shown in the lowest panel of Fig.~4. Similar to model B, Pop II star 
dominate the star formation history at all but the highest redshifts. Indeed, in our approach
the transition between Pop III and Pop II dominated star formation epochs is controlled
only by the strength of chemical feedback ($f_{sn}$) and it is independent of the assumed 
star formation efficiencies (see section \ref{sec:chem}). However, contrary to model B,
the star formation density depends on the escape fraction only indirectly, through 
the strength of radiative feedback. For this reason, the shaded regions
representing the range of allowed star formation densities are not degenerate only 
after reionization which, for $C = 10$  and $f_{esc}=0.5$ occurs
at redshift $z_{\rm rei} = 15$ (for Pop II stars, which are preferentially 
hosted in large halos, this effect is negligible).

\begin{figure}
\centerline{\epsfig{figure=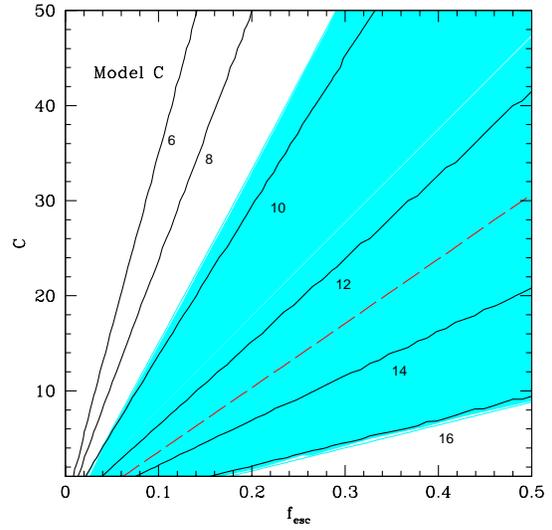, height=8.0cm}}
\caption{\footnotesize Same as Figs.~4 and ~5 but for model C and reionization
  redshifts 6, 8, 10, 12, 14 and 16.}
\end{figure}

The implications for the history of reionization are
summarized in Fig.~7, which shows that model C is in much better agreement
with observations than model A and B. The optical depth observed by
WMAP can be reproduced with $f_{esc} \le 0.15$ even if the clumping
factor is as high as $C \le 25$. 

Thus, when the observed $z \approx 10$ counts are used to set an 
upper limit to the allowed star formation efficiency, Pop~III 
stars forming with a conventional IMF are more efficient 
sources of cosmic reionization than very massive stars. 
Note that this conclusion is independent on issues related to
chemical and radiative feedback. Indeed, both models B and C assume
maximally efficient chemical feedback and radiative feedback plays
a minor role for model C because the predicted $z \approx 10$ counts are
dominated by massive Pop~II galaxies, which are largely unaffected by
photoevaporation. 
\begin{figure}
\centerline{\epsfig{figure=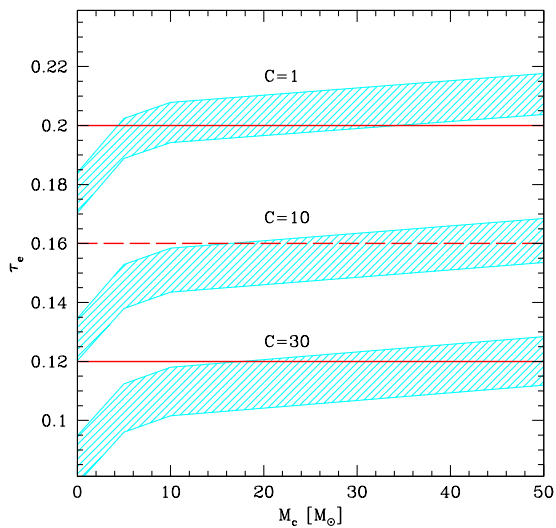, height=8cm}}
\caption{\footnotesize Optical depths as a function of the
characteristic stellar mass of Pop~III stars, $M_c$, assumed to
form with a Larson IMF (see text). The different shaded regions
correspond to $C=1, 10,$ and 30. The emission properties of Pop~III
have been modified according to each assumed value of $M_c$ and 
$f_{esc}=0.1$ has been adopted. For each $C$, the upper and lower
limits to the shaded region corresponds, respectively, to 3 and 
0 $z \approx 10$ galaxies in NICMOS UDFs.}
\end{figure}

Fig.7 shows that the best-fit value of optical depth implied by WMAP
observations, $\tau_e = 0.16$, can only be reproduced by model C 
with $f_{esc} \le 0.15$ if $C \le 8$. 
If larger values of $C$ are to be favored or the observed optical 
depth will be corrected up-ward in future WMAP data releases, 
a revision of model C is to be considered. 

However, unless the NICMOS UDFs counts have under-estimated 
by a large factor the number of $z \approx 10$ sources, our 
results suggest that the only plausible revision of model C is
to modify the primordial IMF by increasing the characteristic 
mass but allowing Pop~III stars to form only in the 
conventional mass range, $1~\Msun \le M_\star \le 100~\Msun$. 
In Fig.~8 we show the predicted optical depth as a function of 
the Pop~III characteristic stellar mass, $M_c$. 
We assume that the characteristic stellar mass enters
in the definition of the stellar IMF as suggested by Larson (1998),

\be
\frac{dN}{d\log M_\star} \propto \left ( 1 + \frac{M_\star}{M_c}\right)^{-1.35},
\ee
\noindent
which, for $M_c=0$ leads to the standard Salpeter IMF. 

For a given $M_c$, we compute $\tau_e$ considering
the proper emission rate of ionizing photons (taken from Table~1 of 
Ciardi \& Ferrara 2005), assuming $f_{esc} = 0.1$, three 
different clumping factors $C=1, 10,$ and 30, and that either all 3 or
none of the candidates identified by Bouwens et al. (2005) in NICMOS
UDFs represent $z \approx 10$ galaxies (upper and lower limits to each 
shaded region). 
Fig.~8 shows that the predicted optical depth depends more on the
clumping factor than on the assumed value of $M_c$. 
If $C=10$, the predicted $\tau_e$ is in agreement with
the observed values even assuming $M_c=0$, consistent with the results
shown in Fig.~7. If $C=30$, WMAP and NICMOS UDFs data can
be reconciled only if all three candidates are confirmed to be
$z \approx 10$ galaxies and if the characteristic Pop~III stellar mass 
is $\ge 20 \,\Msun$. Finally, if the observed $\tau_e$ will be 
corrected up-ward in future analyses of WMAP data, none of the models
considered, independently of the assumed $M_c$, will be able to
reconcile the observed reionization history with the counts at $z \approx
10$, unless $C<10$.

\section{Discussion}

The tentative detections of $z \approx 10$ galaxies in NICMOS UDFs by
Bouwens et al. (2005) provides tight constraints on the relative 
contributions of Pop~III and Pop~II stars to the galaxy luminosity 
function. When coupled to the reionization history of the universe
implied by the WMAP satellite, this collection of data sets represents 
already a very demanding benchmark for galaxy evolution models. 

Stimulated by these data, we have followed the
assembling history of high-redshift galaxies hosting the first stars
in a hierarchical model for structure formation. 
Adopting simple prescriptions for chemical and radiative feedbacks along with
available data, we have constrained the evolution of the IMF of the first stars.
In particular we find that:

\begin{enumerate}
\item Two types of feedbacks need to be considered: 
  (ia) radiative feedback, which suppresses star formation 
  in H$_2$-cooling halos and the formation of low-mass galaxies 
  because of photoionization heating following reionization, and
  (ib) chemical feedback due to the progressive
  enrichment of (Pop~II) star forming gas by the metals released by the 
  first (Pop~III) stellar explosions.
 
\item If chemical feedback is weak ($f_{sn}=0.1$) 
  Pop~III stars can form in relatively large halos and
  dominate the evolution of the cosmic star formation density down to
  redshift $z \lsim 5$. More realistic scenarios are found assuming 
  $f_{sn}=1$, where Pop~III stars are confined to form only in the
  smallest halos at each redshift.

\item The combined effect of radiative and chemical feedbacks fails to
  suppress the formation of massive Pop~III stars at $z \lsim 10$.  
  
\item If Pop~III stars are assumed to form with a Top-Heavy IMF, that
  is with masses $M_\star \ge 100 \, \Msun$, Pop~III galaxy are always
  found to dominate the predicted  $z \approx 10$ NICMOS UDFs counts. 
  Because of their strong Ly$\alpha$ and nebular luminosities, very
  tight upper limits on the star formation efficiency in $T_{vir} \ge
  10^4$~K halos are placed by the paucity of observed counts. The
  corresponding reionization histories fail to reproduce the observed
  WMAP optical depths unless uncomfortably large values for the escape
  fraction of ionizing photons and/or too small values for the IGM 
  clumping factor are assumed.

\item Reionization constraints from WMAP and observed $z \approx 10$
  counts in NICMOS UDFs can be reconciled if Pop~III stars form 
  according to a Larson IMF with stellar masses in the
  conventional range $1 \, \Msun \le M_\star \le 100 \, \Msun$ and
  a characteristic mass $M_c \ge 0$ if the clumping factor is  $\le
  10$, or $M_c \ge 20 \, \Msun$ if the clumping factor is as large as
  30. In the latter case, at least 1 candidate identified by 
  Bouwens et al. (2005) has to be a galaxy at $z \approx 10$.

\end{enumerate}  

Our analysis is based on the assumption that chemical feedback 
propagates along the hierarchy of galaxy mergers from progenitors
to descendants rather than through metal-enriched galaxy outflows 
(Scannapieco et al. 2003). By adopting this simplified scheme, we
are implicitly assuming that (i) even if a fraction of metals is ejected 
out of the host halo, the metals that are kept within the potential 
well are sufficient to pollute the gas within the host system or 
within any unpolluted halo that the system will later merge with, 
to a metallicity level larger than $Z_{\rm cr}$ and that (ii) metal
pollution from neighboring galaxies is negligible.  
In spite of these simplifying assumptions, the resulting 
Pop~III and Pop~II star formation histories are consistent with 
the predictions of Scannapieco et al. (2003). In particular, even
assuming a maximally efficient chemical feedback, the transition is
always predicted to be very smooth, with Pop~III stars which continue
to form in the smallest halos at each redshift down to $z \lsim 5$.
Interestingly enough, our conclusion is independent of the primordial 
IMF and/or star formation efficiency. We show that models where all
Pop~III halos are able to metal pollute their descendants ($f_{sn}=1$)
are favored by observations. The smoothness of the cosmic IMF
transition is therefore strictly linked to the inhomogeneous character
of the metal enrichment process, independently of how it is
implemented in the models. 

Even if we have adopted a very approximate description of the thermal
evolution of the cosmic gas, we find that radiative feedback following
reionization can play an important role in the extinction of Pop~III
stars. In particular, if reionization occurs at redshifts $z \approx 15$
($z \approx 10$) the baryon fraction in $10^8\,\Msun$ halos is reduced 
to 70\% (99\%) of its universal value ($\Omega_B/\Omega_M$) at redshift
10 and to 10\% (30\%) of this value at redshift 5. Thus, the
baryon content in the smallest halos where Pop~III stars
can form at low redshift is severely limited by radiative feedback.
This is strongly related to the epoch of cosmic reionization which,
in turn, depends on the degree of inhomogeneities which characterize
the cosmic gas. We have parametrized this quantity through a constant
clumping factor $C$ and we have explored different values of $C$
ranging from 1 to 50. However, it is likely that the degree of
clumpiness of the cosmic gas varies with time, rising from $C \approx 1$
at high redshift to a few tens at $z \approx 10$ as structures go 
increasingly non-linear. 

Recently, Iliev et al. (2005) have studied the effect
of IGM clumping at redshifts $z \ge 10$, providing a fit to the
evolution of the clumping factor (see their Fig.~7) where $C \approx 2$
at redshift 30 and raises to $C \approx 8$ at redshift 10. Adopting this 
evolution for the clumping factor and $f_{esc}=0.1$ we find that 
models A and B are unable to reproduce the optical depths inferred 
from WMAP data whereas in model C we find that reionization is
complete at redshift 13.5 and that the resulting optical depth is
$\tau_e = 0.1734$ in very good agreement with the observed value. 

Thus, our analysis suggests that Pop~III stars forming with masses in
the same dynamical range as Pop~II/I stars, but possibly with higher
characteristic stellar mass, appear to be favored by observations.
Several scenarios that enable the formation of 
low-mass stars in metal-free environments have been proposed, 
among which the bi-modal IMF (Nakamura \& Umemura 1999, 2002), 
enhanced HD-cooling in relic HII regions (Uehara \& Inutsuka 2000;
Nakamura \& Umemura 2002), the influence of
a strong UV radiation field due to the vicinity of a very massive star
(Omukai \& Yoshii 2003), fragmentation in shock-compressed shells 
induced by the first very massive SN explosions (Mackey, Bromm \& Hernquist
2003; Salvaterra, Ferrara \& Schneider 2004).

However, all these processes either require very special conditions,
generally related to the existence of a previous generation of 
very massive stars, or have failed to be realized in realistic
numerical simulations. 
In spite of this, observations seem to indicate that either 
radiative feedback effects during the accretion phase on protostellar
cores and/or a combination of the above mentioned processes 
can lead to the formation of Pop~III stars with masses $\le 100 \Msun$.
According to the results of our analysis, the latter contribution to
the Pop~III stellar IMF must be dominant at least at redshifts  $z \approx 10$. 
  
If this scenario is correct, relics of the low-mass Pop~III star
formation mode should be still present in our Galaxy,
eventually contributing to the lowermost end of the
metallicity distribution function (Salvadori, Schneider \& Ferrara in prep).

\section*{Acknowledgments}
We acknowledge the use of PINOCCHIO v 1.0 package written by Pierluigi Monaco, Tom Theuns \& Giuliano Taffoni
(http://www.daut.univ.trieste.it/pinocchio). This work has been partially supported by the Research and 
Training Network ``The Physics of the Intergalactic Medium'' set up by
the European Community under the contract HPRN-CT-2000-00126.


\begin{thebibliography}{99}
\bibitem[1]{Abel02}Abel T., Bryan G. L., Norman M. L. 2002, Science, 295, 93
\bibitem[2]{Barkana01}Barkana, R. \& Loeb, A. 2001, PhR, 349, 125
\bibitem[3]{Bouwens04} Bouwens R. J. et al. 2004, , ApJ, 616, L79
\bibitem[4]{Bouwens05} Bouwens R. J., Illingworth G. D., Thomson R. I., \& Franx M, 2005,
ApJ, 624, L5
\bibitem[5]{Bromm01}Bromm, V., Ferrara A., Coppi P. S., Larson R. B. 2001, MNRAS, 328, 969
\bibitem[6]{Bromm01b}Bromm, V., Kudritzki, R. P. \& Loeb, A. 2001, ApJ, 552, 464
\bibitem[7]{Bromm02}Bromm, V., Coppi P. S., Larson R. B. 2002, ApJ, 564, 23
\bibitem[8]{Bruzual}Bruzual, G. \& Charlot, S. 2003, 344, 1000
\bibitem[9]{Ciardi05}Ciardi, B. \& Ferrara, A. 2005, Space Science Reviews, 116, 625
\bibitem[10]{Ciardi03}Ciardi, B., Ferrara, A., White, S. D. M. 2003, MNRAS, 344, L7
\bibitem[11]{Cen03}Cen, R. 2003, ApJ, 591, 12
\bibitem[12]{Furlanetto05}Furlanetto, S. \& Loeb, A. 2005, ApJ in
  press, (astro-ph/0409736)
\bibitem[13]Gnedin, N., Ostriker, J. 1997, ApJ, 474, 581
\bibitem[14]{Gnedin}Gnedin, N. 2000, ApJ, 542, 535
\bibitem[15]{Haiman03}Haiman, Z. \& Holder, G. P. 2003, ApJ, 595, 1
\bibitem[16]{Heger02}Heger, A. \& Woosley, S. 2002, ApJ, 567, 532
\bibitem[17]{Iliev05}Iliev, I., Scannapieco, E., Shapiro, P. 2005, ApJ, 624, 491
\bibitem[17]{Kogut03}Kogut, A. et al. 2003, ApJS, 148, 161
\bibitem[18]{Larson98}Larson, R. B. 1998, MNRAS, 301, 569
\bibitem[19]{Mackey}Mackey, J., Bromm, V., Hernquist, L. 2003, ApJ, 586, 1
\bibitem[20]{Madau99}Madau, P., Haardt, F., Rees, M.~J. 1999, ApJ, 514, 648
\bibitem[21]{Monaco02}Monaco, P., Theuns, T. \& Taffoni, G. 2002, MNRAS, 331, 587
\bibitem[22]{Nakamura}Nakamura, F., Umemura, M. 1999, ApJ, 515, 239
\bibitem[23]{Nakamura}Nakamura, F., Umemura, M. 2002, ApJ, 569, 549
\bibitem[24]{Omukai98}Omukai, K. \& Nishi, R. 1998, ApJ, 508, 141
\bibitem[25]{Omukai00}Omukai, K. 2000, ApJ, 534, 809
\bibitem[26]{Omukai03a}Omukai, K. \& Palla, F. 2003, ApJ, 589, 677
\bibitem[27]{Omukai03b}Omukai, K. \& Yoshii, Y. 2003, ApJ, 599, 746
\bibitem[28]{Omukai05}Omukai, K., Tsuribe, T., Schneider, R., Ferrara, A. 2005, ApJ, 626, 627
\bibitem[29]{Ricotti04}Ricotti, M. \& Ostriker, J. P., 2004, MNRAS, 350, 539
\bibitem[30]{Ripamonti02}Ripamonti, E., Haardt, F., Ferrara, A., \& Colpi, M., 2002, MNRAS, 334, 401
\bibitem[31]{Salvaterra03b}Salvaterra, R., Ferrara A., Schneider R. 2004, New Astronomy, 10, 11
\bibitem[31]{Salvaterra03b}Salvaterra, R. \& Ferrara A. 2003, MNRAS, 339, 973
\bibitem[32]{Scannapieco03}Scannapieco, E., Schneider, R. \& Ferrara, A. 2003, ApJ, 589, 1
\bibitem[33]{Scannapieco05}Scannapieco, E., Madau, P., Woosley, S.,
  Heger, A., Ferrara, A. 2005, ApJ, in press (astro-ph/0507182)   
\bibitem[34]{Schaerer02}Schaerer, D. 2002, A\&A, 382, 28
\bibitem[35]{Schneider02}Schneider R., Ferrara A., Natarajan P., Omukai K. 2002, ApJ, 571, 30
\bibitem[36]{Schneider03}Schneider R., Ferrara A., Salvaterra R., Omukai K., Bromm V. 2003, Nature, 422, 869
\bibitem[37]{Sigward05}Sigward, F., Ferrara, A. \& Scannapieco, E. 2005, MNRAS, 358, 755
\bibitem[37]{Sokasian04}Sokasian, A., Yoshida, N., Abel, T.,
  Hernquist, L., Springel, V. 2004, MNRAS 350, 47
\bibitem[38]{Spergel03}Spergel, D. L. et al. 2003, ApJS, 148, 175
\bibitem[39]{Stahler86}Stahler, S. W., Palla, F., \& Salpeter,
  E. E. 1986, ApJ, 302, 590
\bibitem[40]{Tan04}Tan, J. C. \& McKee, C. F. 2004, ApJ, 603, 383
\bibitem[41]{Taffoni02}Taffoni, G., Monaco, P. \& Theuns, T. 2002, MNRAS, 333, 623
\bibitem[42] Uehara, H. \& Inutsuka, S. 2000, ApJ, 531, L91
\bibitem[43]{Venkatesan03}Venkatesan, A., Tumlinson, J., Shull,
  J.~M. 2003, ApJ, 584, 621  
\bibitem[44]{Wyithe03}Wyithe, J. S. B. \& Loeb, A. 2003, ApJ, 588, L69
\bibitem[45]{Yoshii80}Yoshii, Y., Sabano, Y.  1980, PASJ, 32, 229  
\end{thebibliography}
\end{document}